\documentclass[a4paper, 11pt]{article}
\usepackage{graphicx}
\usepackage{jheppub}
\usepackage[utf8]{inputenc}
\usepackage{caption}
\usepackage{subcaption}
\usepackage{hyperref}
\usepackage{enumerate}
\usepackage{wrapfig}
\usepackage{amsmath}
\usepackage{amsfonts}
\usepackage{booktabs}
\usepackage{comment}
\usepackage{cancel}
\usepackage{vmargin}
\setmargins{2.5cm}{1.5cm}{16.5cm}{23.42cm}{10pt}{1cm}{0pt}{2cm}
\usepackage{multirow}
\usepackage{slashed}
\usepackage{braket}
\usepackage{amssymb}
\usepackage{setspace}
\usepackage{amssymb}
\usepackage{url}
\usepackage{xcolor}
\usepackage{xspace}
\usepackage{cleveref}
\usepackage{mathabx}
\usepackage{tikz}
\usepackage{tikz-feynman}
\tikzfeynmanset{compat=1.1.0}
\notoc
\title{Non-leptonic two-body $\boldsymbol{B}$ decays}
\author[a,b]{M. Burgos Marcos}
\date{Maastricht University}

\affiliation[a]{Gravitational Waves and Fundamental Physics (GWFP), Maastricht University, Duboisdomein 30, NL-6229 GT Maastricht, the Netherlands}
\affiliation[b]{Nikhef, Science Park 105, NL-1098 XG Amsterdam, the Netherlands}

\abstract{Hadronic $B$ meson decays offer a rich laboratory to study the interplay of the weak and the strong interaction. However, they remain among the most challenging processes to describe as non-perturbative strong-interaction effects enter in both the initial and final states. Their theoretical description therefore relies on approximations and theoretical models. With the increasingly precise data from LHCb and Belle II, the question of how well these models capture the dynamics of the decays is not only pressing, but can finally be answered quantitatively. In this contribution, I review two approaches to this question: QCD Factorization (QCDF) and a data-driven approach based on flavour symmetries, which provides a way to test it. QCDF provides a systematic framework to compute the hadronic topologies by exploiting the large mass of the $b$ quark. Its predictions are, however, limited by endpoint divergences in the hard spectator-scattering and annihilation contributions. In addition, tensions with data have been observed in decays such as $B\to K\pi$ and $B\to Dh$, motivating a more thorough test of QCDF against data. Flavour symmetries provide the tool for this test: by relating the topological amplitudes of different decay channels, they allow these amplitudes to be constrained directly from data through global fits. These fits not only describe the observables for which experimental data is available but also provide predictions for those yet to be measured. Finally, I confront the QCDF predictions with the results of these global fits, assessing where this comparison is currently meaningful and which measurements would most improve it.}
\begin{document}
\maketitle
\newpage
\setcounter{page}{1}
\spacing{1.35}

\section{Introduction}

Decays of $B$ mesons are an ideal testing ground for the Standard Model (SM): they probe the weak interaction, through CP violation and the structure of the CKM matrix, and the strong interaction, through the dynamics that bind quarks into hadrons. $B$ meson decays are conventionally classified into leptonic, semileptonic and hadronic, according to the final state produced by the weak decay of the $b$ quark. This classification also reflects an increasing degree of theoretical complexity. Leptonic decays are governed by a single hadronic parameter, the decay constant. Semileptonic decays involve form factors that can be computed or constrained via lattice QCD \cite{FermilabLattice:2015mwy,Flynn:2015mha,Colquhoun:2022atw,Bouchard:2014ypa} and light-cone sum rules \cite{Gubernari:2018wyi, Melic:2025uha,Leljak:2021vte}. Hadronic decays, in contrast, are sensitive to non-perturbative strong-interaction effects not only in the $B$-to-meson transition and in the hadronization of the additional final-state meson, but also in the interactions linking the two. These long-distance QCD effects make perturbative calculations considerably more involved than in the other two categories, and therefore the theoretical description of hadronic $B$ decays has remained comparatively limited in precision.

The experimental landscape of hadronic $B$ decays has evolved considerably over the past two decades. The large samples collected by collider experiments and $B$ factories, such as LHCb \cite{LHCb:2008vvz} and Belle II \cite{Belle-II:2018jsg}, have provided an increasingly precise set of branching ratio and CP asymmetry measurements for hadronic modes. At this level of precision, the data can be used to test theoretical frameworks quantitatively. This has renewed the interest in the approaches used to compute hadronic $B$ decay amplitudes, among them perturbative QCD (pQCD) \cite{Keum:2000ph,Keum:2000wi,Lu:2000em}, Factorization-Assisted Topologies (FAT) \cite{Li:2012cfa,Qin:2013tje,Zhou:2015jba}, QCD Factorization (QCDF) \cite{Beneke:1999br,Beneke:2000ry,Beneke:2001ev,Beneke:2003zv} and phenomenological analyses based on $SU(3)_F$ flavour symmetry \cite{BurgosMarcos:2025xja,Huber:2021cgk,Berthiaume:2023kmp,Bhattacharya:2025wcq,Fang:2026fhl}. Here I will focus on the latter two, which are closely connected since data-driven $SU(3)_F$ analyses offer a way to test QCDF directly against experiment. 

This contribution is organized as follows. Sec.~\ref{sec:QCDF} gives a brief overview of QCD Factorization, its general setup and open challenges. Sec.~\ref{sec:pheno} presents data-driven approaches based on flavour symmetries, focusing on $SU(3)_F$ symmetry with factorizable corrections. The comparison between these global fits and QCDF is discussed in Sec.~\ref{sec:QCDFvsPheno}. I conclude in Sec.~\ref{sec:conclusions}.

\subsection{General description of hadronic $B \to M_1M_2$ decays}\label{sec:generaldesc}

The amplitude of any two-body hadronic $B$ meson decay can be written, in full generality, as:
\begin{equation}
    \mathcal{A}(\bar B_q\to M_1M_2)=i\frac{G_F}{\sqrt{2}}\left[\lambda_u^{(p)}\mathcal{T}+\lambda_c^{(p)}\mathcal{P}\right]
    \label{eq:ampgeneral}
\end{equation}
using the CKM unitarity relation $\lambda_u^{(p)}+\lambda_c^{(p)}+\lambda_t^{(p)}=0$, with $\lambda_i^{(p)}\equiv V_{ib}V_{ip}^*$ and $p=d,s$.

The amplitudes $\mathcal{T}$ and $\mathcal{P}$ encode the contributions from several different topologies. In particular, the topologies contributing to $\mathcal{T}$ can be obtained via \cite{He:2018php,Huber:2021cgk}:
\begin{equation}
\begin{aligned}
 {\cal T} =&\, T ~B_i (M)^{i}_j   \bar H^{jl}_k  (M)^k_l   + C~B_i (M)^{i}_j \bar H^{lj}_k  (M)^k_l +  A~B_i \bar H^{il}_j   (M)^j_k (M)^{k}_l \\
 &+  E~B_i  \bar H^{li}_j (M)^j_k (M)^{k}_l + {T_{P}} B_i (M)^{i}_j   (M)^j_k \bar H^{lk}_{l}  + {T_{PA}} B_i \bar H^{li}_{l}  (M)^j_k (M)^{k}_j ,
 \end{aligned}
 \label{eq:Tformula}
\end{equation}
with $\bar H_1^{12}=\delta_{pd}, \, \bar H_1^{13}=\delta_{ps}$, the light pseudoscalar meson matrix
\begin{equation}
\label{eq:Mmatrix}
\begin{aligned}
   M &= \left( \begin{array}{ccc} 
    \frac{\pi^0}{\sqrt 2} + \frac{\eta_8}{\sqrt 6}
     & \pi^- & K^- \\
    \pi^+ & - \frac{\pi^0}{\sqrt2} + \frac{\eta_8}{\sqrt 6}
      & \bar K^0 \\
    K^+ & K^0 & -2 \frac{\eta_8}{\sqrt 6}  \end{array} \right)
\end{aligned}
+ 
\begin{aligned}
   \left( \begin{array}{ccc} 
   \frac{\eta_0}{\sqrt 3}
     & 0 & 0 \\
    0 & \frac{\eta_0}{\sqrt 3}
      & 0 \\
    0 & 0 & \frac{\eta_0}{\sqrt 3}  \end{array} \right),
\end{aligned}
\end{equation}
and the $B$ meson vector $B_i=(B^+,B^0,B^0_s)$. The topologies contributing to $\mathcal{P}$ can be obtained by applying the following substitutions to Eq.~\eqref{eq:Tformula} :
\begin{align}
T&\to P_{T}, & C & \to P_{C}, &
A&\to P_{TA},& \nonumber\\
T_{P}&\to P, &
E&\to P_{TE}, & T_{PA}&\to P_{A}.
\label{eq:TPrelationTop}
\end{align}

Because the $\mathcal{T}$ and $\mathcal{P}$ amplitudes are present simultaneously in all $B \to M_1M_2$ decays and carry different weak phases, all these decays are subject to CP violation within the SM. 

It is important to stress that this decomposition follows directly from the structure of the effective weak Hamiltonian and the unitarity of the CKM matrix, and is therefore completely general: no theoretical assumptions have been imposed so far. Assumptions enter only when the topological amplitudes are evaluated. In a theoretical calculation, an approximation must be chosen that exploits the hierarchy of scales of the decay, as in the heavy-quark expansion of QCDF discussed in Sec.~\ref{sec:QCDF}. In a data-driven analysis, amplitudes of different decay channels must be related through additional input, such as the flavour symmetries discussed in Sec.~\ref{sec:pheno}.

\section{Overview of QCD Factorization}\label{sec:QCDF}

The topological amplitudes introduced in the previous section cannot be calculated from first principles without additional theoretical assumptions. QCD Factorization (QCDF) provides a theoretical framework in which these topologies can be systematically calculated order by order in perturbation theory \cite{Beneke:1999br,Beneke:2000ry,Beneke:2001ev,Beneke:2003zv}. It exploits the hierarchy $m_b\gg \Lambda_{\mathrm{QCD}}$ to expand the decay amplitude in powers of $\alpha_s$ and $\Lambda_\mathrm{QCD}/m_b$. In the heavy-quark limit, and at leading power in $\Lambda_\mathrm{QCD}/m_b$, the non-perturbative content of the decay, namely the $B \to M_1$ form factor and the mesons' Light-Cone Distribution Amplitudes (LCDAs), is factorized from the perturbative term, which encodes the short-distance dynamics of each topology. Schematically, the amplitude of each topology is then the product of a non-perturbative prefactor, built from the form factor and decay constants, and a perturbative coefficient. QCDF distinguishes two classes of topologies, tree-like and annihilation-like, whose perturbative coefficients are denoted $a_i$ and $b_i$, respectively.
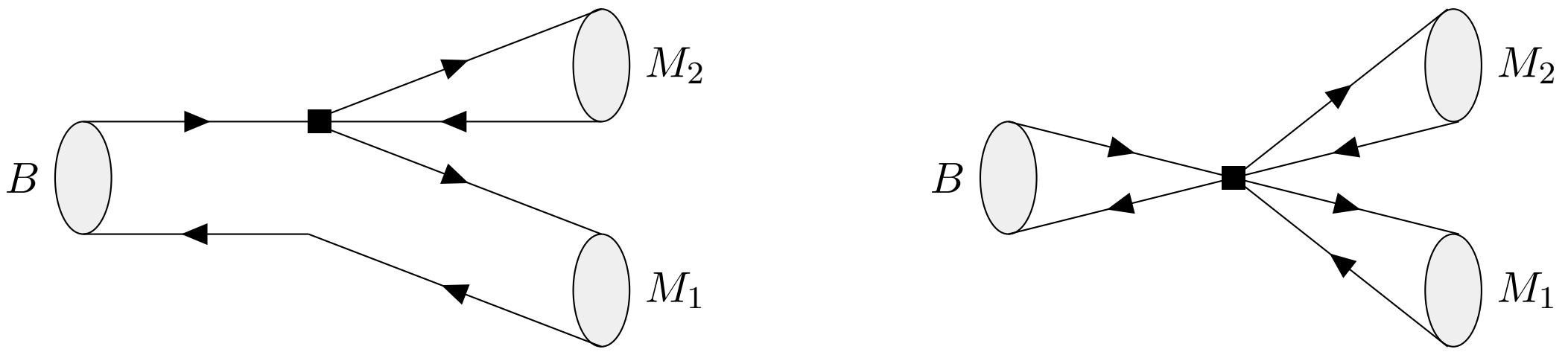
\begin{figure}[t]
\begin{subfigure}[b]{0.49\textwidth}
    \begin{tikzpicture}
        \begin{feynman}
    \vertex (a) ;
    \vertex [dot, minimum size=6pt] at ($(a) + (2cm, 0 cm)$) (b);
    \vertex at ($(b) + (2.6cm, 1 cm)$) (c);
    \vertex at ($(a) + (0cm, -1cm)$)(d) ;
    \vertex at ($(d) + (2cm, 0cm)$)(e) ;
    \vertex at ($(e) + (2.6 cm, -1cm)$) (f);
    \vertex at ($(b) + (1.1 cm, -0.5cm)$) (g);
    \vertex at ($(g) + (1.5 cm, 0.5cm)$) (h);
    \vertex at ($(g) + (1.5 cm, -0.5cm)$) (i);
    \node[draw,ellipse,minimum width=1cm,minimum height=0.5cm, rotate=90,fill=gray!30,
  fill opacity=0.4] (Bc) at ($(a)!0.5!(d)$) {};
  \node[draw,ellipse,minimum width=1cm,minimum height=0.5cm, rotate=90,fill=gray!30,
  fill opacity=0.4] (Bp) at ($(i)!0.5!(f)$) {};
  \node[draw,ellipse,minimum width=1cm,minimum height=0.5cm, rotate=90,fill=gray!30,
  fill opacity=0.4] (pi0) at ($(c)!0.5!(h)$) {};
  \node[anchor=east, xshift=-0.25cm] at (Bc.center) {$B$};
  \node[anchor=west, xshift=0.25cm] at (Bp.center) {$M_1$};
  \node[anchor=west, xshift=0.25cm] at (pi0.center) {$M_2$};
  \node[rectangle, fill=black, inner sep=0pt, minimum size=6pt] at ($(b) + (0.1cm, 0cm)$);
    
    \diagram* {
      (a) -- [fermion] (b);
      (b) -- [fermion] (c),
      (e) -- [fermion] (d);
      (f) -- [fermion] (e);
      (h) -- [fermion] (b);
      (b) -- [fermion] (i);
    };
  \end{feynman}
    \end{tikzpicture}
    \caption{$\mathcal{T}^\mathrm{QCDF}$}
    \label{fig:TQCDF}
\end{subfigure}
\begin{subfigure}[b]{0.49\textwidth}
    \begin{tikzpicture}
        \begin{feynman}
    \vertex (a) ;
    \vertex at ($(a) + (0cm, -1 cm)$) (b);
    \vertex at ($(b) + (2cm, 0.5 cm)$) (c);
    \vertex at ($(c) + (1.9cm, 1.5cm)$)(d) ;
    \vertex at ($(d) + (0.1cm, -1.cm)$)(e) ;
    \vertex at ($(e) + (0 cm, -1cm)$) (f);
    \vertex at ($(f) + (-0.1 cm, -1.cm)$) (g);
    \node[draw,ellipse,minimum width=1cm,minimum height=0.5cm, rotate=90,fill=gray!30,
  fill opacity=0.4] (Bc) at ($(a)!0.5!(b)$) {};
  \node[draw,ellipse,minimum width=1cm,minimum height=0.5cm, rotate=90,fill=gray!30,
  fill opacity=0.4] (Bp) at ($(d)!0.5!(e)$) {};
  \node[draw,ellipse,minimum width=1cm,minimum height=0.5cm, rotate=90,fill=gray!30,
  fill opacity=0.4] (pi0) at ($(f)!0.5!(g)$) {};
  \node[anchor=east, xshift=-0.25cm] at (Bc.center) {$B$};
  \node[anchor=west, xshift=0.25cm] at (Bp.center) {$M_2$};
  \node[anchor=west, xshift=0.25cm] at (pi0.center) {$M_1$};
  \node[rectangle, fill=black, inner sep=0pt, minimum size=6pt] at (c);
    
    \diagram* {
      (a) -- [fermion] (c);
      (c) -- [fermion] (b),
      (c) -- [fermion] (d);
      (e) -- [fermion] (c);
      (c) -- [fermion] (f);
      (g) -- [fermion] (c);
    };
  \end{feynman}
    \end{tikzpicture}
    \caption{$\mathcal{A}^\mathrm{QCDF}$}
    \label{fig:AQCDF}
\end{subfigure}
\caption{Schematic representation of tree-like and annihilation-like topologies contributing to non-leptonic $B$-meson decays within QCD Factorization.}
\end{figure}

The tree-like amplitudes $\mathcal{T}^\mathrm{QCDF}$ represented in Fig.~\ref{fig:TQCDF} are given by:
\begin{equation}
    \mathcal{T}^\mathrm{QCDF}_{B\to M_1M_2} = M_B^2 F_0^{B\to M_1} (q^2) f_{M_2} a_i^p(M_1M_2)
\end{equation}
where $F_0^{B\to M_1}$ represents the $B\to M_1$ form factor, $f_M$ is the decay constant of the meson $M$ and the perturbative contribution at NLO is given by \cite{Beneke:2003zv}:
\begin{equation}
 a^p_i(M_1M_2)=\left(C_i + \frac{C_{i\pm 1}}{N_c}\right)N_i(M_2) + \frac{C_{i\pm 1}}{N_c} \frac{C_F\alpha_s}{4\pi}\left[V_i(M_2) + \frac{4\pi^2}{N_c}H_i(M_1M_2)\right]+P_i^p(M_2).
 \label{eq:QCDFtree}
\end{equation}
The first term corresponds to the leading order contribution, $V_i(M_2)$ contains the NLO vertex corrections, $H_i(M_1M_2)$ represents the hard spectator scattering contribution, and $P_i^p(M_2)$ are the penguin contractions. 
While the $a_i$ coefficients are given at NLO in Eq.~\eqref{eq:QCDFtree}, NNLO calculations are available for a subset of topologies. The current NNLO results for the colour-allowed tree ($\alpha_1$ or $a_1$), colour-suppressed tree ($\alpha_2$ or $a_2$) \cite{Beneke:2009ek} and QCD penguin ($a_4^p$) coefficients \cite{Bell:2020qus} are:
\begin{equation}
    \begin{aligned}
        \alpha_1(\pi\pi) &= 1.000^{+0.029}_{-0.069} + 0.011^{+0.023}_{-0.050}i, \\
        \alpha_2(\pi\pi) &= 0.240^{+0.217}_{-0.125} - 0.077^{+0.115}_{-0.078}i, \\
        a_4^u(\pi \bar K)/10^{-2} &= -2.12^{+0.48}_{-0.29}-1.56 ^{+0.29}_{-0.15}i,
        \label{eq:QCDFresults}
    \end{aligned}
    \end{equation}
 where the larger uncertainties in $\alpha_2$ relative to its colour-allowed counterpart $\alpha_1$ are indicative of its greater sensitivity to perturbative corrections.
 
Weak annihilation amplitudes, represented in Fig.~\ref{fig:AQCDF}, are analogously written as:
\begin{equation}
    \mathcal{A}^\mathrm{QCDF}_{B\to M_1M_2} = f_Bf_{M_1}f_{M_2} b_i(M_1M_2)
\end{equation}
where, for example, the $b_1$ coefficient is given by
\begin{equation}
 b_1=\frac{C_F}{N_c^2}C_1\pi\alpha_s \int_0^1 dxdy\left\{ \Phi_{M_2}(x)\Phi_{M_1}(y) \left[\frac{1}{y(1-x\bar y)} + \frac{1}{\bar x^2 y} \right] + r_\chi^{M_1}r_\chi^{M_2}\Phi_{m_2}(x)\Phi_{m_1}\frac{2}{\bar xy} \right\}
 \label{eq:QCDFann}
\end{equation}
where $\bar x = 1-x$ and analogously for $y$, $r_\chi^M$ are chirally enhanced mass ratios and $\Phi_M$ and $\Phi_m$ are the $M(m)$-meson leading-twist and twist-3 LCDAs respectively. The remaining annihilation topologies can be found in \cite{Beneke:2003zv}.

Note that the distinction between tree-like ($\mathcal{T}^\mathrm{QCDF}$) and annihilation-like ($\mathcal{A}^\mathrm{QCDF}$) amplitudes used in QCDF differs from the $\mathcal{T}/\mathcal{P}$ decomposition used in Eq.~\eqref{eq:ampgeneral}. While the latter is defined by the CKM structure multiplying each amplitude, the QCDF one reflects the way in which factorization is applied: through the $B\to M_1$ form factor and the decay constant of the emitted meson for tree-like topologies, or through the decay constants of all three mesons for annihilation-like ones.

\subsection{Open challenges in QCD Factorization} \label{sec:challengeQCDF}
While QCDF provides a systematic and predictive framework for hadronic decays, several open challenges remain.
\begin{itemize}
    \item \textbf{Endpoint divergences:} the hard spectator scattering term $H_i(M_1M_2)$ within $a_i$ and the full annihilation topologies $b_i$ involve convolutions over soft gluon momentum fractions that diverge at the endpoints of the integration region. These divergences reflect the breakdown of perturbative factorization for soft gluon exchange in these particular topologies, and are not calculable from first principles within QCDF. They are instead modelled as \cite{Beneke:2003zv}:
    \begin{equation}
        X_{H,A}^M =\int_0^1 \frac{dy}{y} \qquad -\frac{1}{2}(X_{H,A}^M)^2=\int_0^1\frac{\ln y}{y}dy,
        \label{eq:endp_div}
    \end{equation}
    introducing an irreducible source of theoretical uncertainty in any decay channel where hard spectator scattering or annihilation contributions are present.
    \item \textbf{Deviations from experimental data:} two channels illustrate the tension between QCDF predictions and experimental measurements. The sum rule in $B \to K\pi$ relates the direct CP asymmetries $\delta(K\pi)\equiv \mathcal{A}_\mathrm{CP}^\mathrm{dir}(K^+\pi^0)-\mathcal{A}_\mathrm{CP}^\mathrm{dir}(K^+\pi^-)$. The experimental results \cite{PDG2025} and QCDF predictions \cite{Bell:2015koa},
    \begin{equation}
        \delta(K\pi)^\mathrm{exp}=-11.0\pm 1.2 \quad \delta(K\pi)^\mathrm{QCDF,2015}=-2.07\pm0.39^{+4.55}_{-2.76}
    \end{equation}
    are found to be in $>3\sigma$ tension, a discrepancy known as the $B\to K\pi$ puzzle. A second, more systematic set of tensions appears in the $B \to Dh$ system. A clear example is $B^0 \to D^-K^+$, which is expected to be well described within QCDF. At leading order in $\Lambda_\mathrm{QCD}/m_b$ it receives the contribution from only a colour-allowed tree, which is expected to be relatively stable under perturbative corrections. Additionally, no annihilation or hard-spectator scattering terms contribute at this order, and hence no endpoint divergences arise. Nevertheless, the branching ratio measurement \cite{PDG2025} and QCDF prediction \cite{Bordone:2020gao} differ significantly:
\begin{equation}
        \mathcal{B}(B^0\to D^-K^+)|_\mathrm{exp}=(2.05\pm0.08)\times 10^{-4} \qquad \mathcal{B}(B^0\to D^-K^+)|_\mathrm{QCDF}=(3.26 \pm 0.15)\times 10^{-4}.
    \end{equation}
Similar discrepancies arise in other $B\to Dh$ channels, ranging from $\sim3\sigma$ to $>6\sigma$ \cite{Huber:2016xod}. Some of these channels can also receive contributions affected by  the endpoint divergences discussed above, making their interpretation less clean. 

The size of these tensions, even in channels under relatively good theoretical control, motivates a more detailed, data-driven assessment of QCDF in this system.
 
     \item \textbf{NNLO calculations only for a subset of topologies:} NNLO calculations are technically challenging and have  only been achieved for the colour-allowed tree, colour-suppressed tree and QCD penguin coefficients quoted in Eq.~\eqref{eq:QCDFresults}. While these results provide valuable control over the perturbative expansion, the lack of NNLO calculations for the remaining coefficients introduces an additional source of uncertainty when assessing the theoretical precision of QCDF.

     \item \textbf{Power corrections:} QCDF is formulated at leading power in $\Lambda_\mathrm{QCD}/m_b$, so higher-power corrections are not systematically included. Unlike the perturbative corrections, these carry no factor of $\alpha_s$, and may therefore be as large as the computable NLO terms. This is the case for annihilation topologies and for chirally enhanced terms such as $r_\chi^M$ \cite{Beneke:2000pw}, which are formally $1/m_b$ suppressed but numerically large. The hard spectator-scattering term, whose short-distance scale is $(m_b\Lambda_\mathrm{QCD})^{1/2}$ rather than $m_b$, is particularly sensitive to them. These effects are connected to the endpoint divergences discussed above and add a further source of theoretical uncertainty.
\end{itemize}

The tensions and limitations discussed above motivate the use of experimental data as a probe of QCDF, both to constrain its otherwise unpredictable endpoint parameters and to identify the origin of the observed discrepancies. This is one of the main goals of the phenomenological studies discussed in the following section.

\section{Flavour symmetries and phenomenological studies}\label{sec:pheno}

The framework introduced in Sec.~\ref{sec:generaldesc} is completely general but cannot be applied directly to phenomenological studies, since multiple topological amplitudes contribute to each decay channel and cannot be determined individually from the available observables. A way to overcome this limitation is to exploit relations between different decay channels provided by flavour symmetries of the strong interaction. These symmetries allow for the simultaneous determination of these topological amplitudes from a combined fit to multiple modes.

A natural starting point is isospin symmetry, which treats the $u$ and $d$ quarks as mass-degenerate and relates, for example, the three $B\to \pi\pi$ channels \cite{Gronau:1990ka,Gronau:2005pq,Fleischer:2018bld}. The analogous symmetry between $d$ and $s$ quarks is known as U-spin \cite{Gronau:2000zy,Gronau:2013mda,Fleischer:1999pa,Fleischer:2018bld}. Both allow a theoretically clean extraction of relative hadronic parameters, without assumptions on the size of individual topologies. However, since they relate only two quark flavours, the number of channels that can be treated simultaneously is limited and, in general, not enough to determine topologies individually. This limitation motivates the extension to $SU(3)_F$ symmetry.

\subsection{$SU(3)_F$ symmetry and factorizable effects}\label{sec:SU(3)sym}

Extending the flavour symmetry to include the three lightest quarks simultaneously leads to $SU(3)_F$ symmetry, under which all sixteen $B\to M_1M_2$ decay modes are related, with $B \in \{B^+,B^0,B^0_s\}$ and $M\in\{\pi,K\}$. Although the $\eta_8$ octet component would also be related under exact $SU(3)_F$ symmetry, the physical $\eta$ and $\eta'$ states receive an additional contribution beyond $SU(3)_F$ symmetry from the $\eta_0$ singlet due to mixing. For this reason, decays involving $\eta$ and $\eta'$ mesons are sometimes treated separately or excluded from the set of channels related by $SU(3)_F$.

Tests of full $SU(3)_F$ symmetry have been performed in two main ways. The first one relies on SM-informed approximations, neglecting the Wilson coefficients $C_7$ and $C_8$ of the electroweak (EW) penguin operators, which are small in the SM. This allows one to relate EW penguins to tree topologies, reducing the total number of independent hadronic parameters \cite{Bhattacharya:2025wcq,Berthiaume:2023kmp}. The second, agnostic approach makes no assumption on the relative size of the topological or Wilson coefficients \cite{BurgosMarcos:2025xja}. Both implementations, however, fail to describe the current experimental data available in $B\to M_1M_2$, indicating that full $SU(3)_F$ symmetry is not a good approximation in these hadronic processes. While this is not necessarily unexpected given the $20-30\%$ $SU(3)_F$-breaking effects generally expected in hadronic processes, the current experimental precision allows one to test whether breaking at this level provides a good description of the data. 

This motivates going beyond the exact $SU(3)_F$ limit. $SU(3)_F$-breaking effects can be incorporated via factorizable corrections, inspired by the structure of QCDF amplitudes discussed in Sec.~\ref{sec:QCDF}, on top of the exact $SU(3)_F$ limit. The first global analysis including factorizable $SU(3)_F$-breaking effects was performed in \cite{BurgosMarcos:2025xja}, and subsequently extended in \cite{Fang:2026fhl} to include $\eta$ and $\eta'$ mesons in the final state.

For tree-like and annihilation-like amplitudes, these factorizable effects are included as the prefactors $A_{M_1M_2}^{B_q}$ and $B_{M_1M_2}^{B_q}$ respectively, which are given by:
\begin{equation}
    A_{M_1M_2}^{B_q} = M_B^2 F_0^{B\to M_1}(m_{M_2}^2)f_{M_2} \qquad B_{M_1M_2}^{B_q}=f_{B_q}f_{M_1}f_{M_2}
\end{equation}
where $M_B$ is the mass of the decaying $B$ meson. By applying these factorizable effects, the $SU(3)_F$ symmetric topologies presented in Eqs.~\eqref{eq:Tformula} and~\eqref{eq:TPrelationTop} become:
\begin{align}
T&= A^{B_q}_{M_1M_2} \tilde\alpha_1, & C & = A^{B_q}_{M_1M_2} \tilde\alpha_2, & T_{P}&= A^{B_q}_{M_1M_2} \tilde\alpha_4^u,
\nonumber \\ 
A&= B^{B_q}_{M_1M_2} \tilde b_2, & E&= B^{B_q}_{M_1M_2} \tilde b_1, &
T_{PA}&= B^{B_q}_{M_1M_2} \tilde b_4^u, 
\label{eq:TopQCDFrelationTree}
\end{align}
and the penguin parameters
\begin{align}
P_T&= \frac{3}{2}~A^{B_q}_{M_1M_2} \alpha_{4,EW}^c, & P_C & =\frac{3}{2} A^{B_q}_{M_1M_2} \alpha_{3,EW}^c, &
P&= A^{B_q}_{M_1M_2} \tilde\alpha_4^c, \nonumber\\  P_{TA}&= \frac{3}{2}B^{B_q}_{M_1M_2} b_{3,EW}^c, 
& P_{TE}&= \frac{3}{2}B^{B_q}_{M_1M_2} b_{4,EW}^c,&
P_A&= B^{B_q}_{M_1M_2} \tilde b_4^c.
\label{eq:TopQCDFrelationPeng}
\end{align}
The $\alpha_i$ and $b_i$ coefficients are $SU(3)_F$-symmetric and fitted to data.

Both analyses including factorizable corrections lead to an excellent description of the data \cite{BurgosMarcos:2025xja,Fang:2026fhl}, indicating that $SU(3)_F$-breaking effects at the order of $20-30\%$ are sufficient to describe these decays.

An important feature of these global fits is that they are not limited to describing observables for which experimental data is available, but they can also predict unmeasured observables. An example is $\mathcal{A}_\mathrm{CP}^\mathrm{mix}(B^0\to\pi^0\pi^0)$:
\begin{equation}
    \mathcal{A}_\mathrm{CP}^\mathrm{mix} (B^0\to\pi^0\pi^0)\equiv \frac{2\mathrm{Im}(\xi_{\pi^0\pi^0})}{1+|\xi_{\pi^0\pi^0}|^2} \quad \mathrm{with}\quad \xi_{\pi^0\pi^0}=-e^{-i\phi_d}\frac{\mathcal{A}(\bar B^0\to\pi^0\pi^0)}{\mathcal{A}(B^0\to\pi^0\pi^0)},
\end{equation}
which had not been measured yet at the time of the analysis in \cite{BurgosMarcos:2025xja}. The global fit predicted:
\begin{equation}
    \mathcal{A}_\mathrm{CP}^\mathrm{mix}(B^0\to\pi^0\pi^0)|_\mathrm{fit}=(-92.3^{+9.3}_{-4.9})\%.
\end{equation}
This observable has since been measured by Belle II \cite{Belle-II:2026kzq},
\begin{equation}
    \mathcal{A}_\mathrm{CP}^\mathrm{mix}(B^0\to\pi^0\pi^0)|_\mathrm{Belle-II}=(-61^{+79}_{-75}\pm 11)\%.
\end{equation}
which is in agreement with the previous prediction, although the current experimental uncertainty is still relatively large.

These fits also identify which observables would most benefit from improved experimental precision. Examples of particular interest are $B^0\to K^0\bar K^0$ and $B_s^0\to K^0\bar K^0$, whose improved precision would help reduce the overall uncertainty on the fit. The specific role played by these channels is discussed in more detail in the following section.

\section{Confronting QCDF with phenomenological fits}\label{sec:QCDFvsPheno}
\begin{figure}[t]
    \centering
    \includegraphics[width=0.6\textwidth]{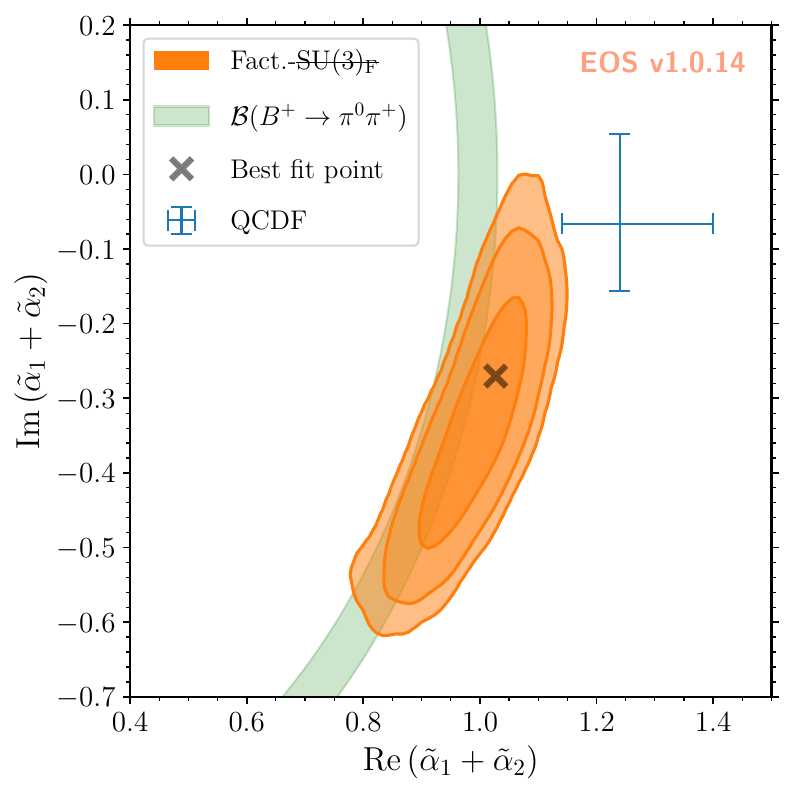}
    \caption{
        Posterior distribution of $\tilde \alpha_1 + \tilde\alpha_2$, showing the $68\%, \, 95\%$ and $99\%$ cumulative contours under the assumption $\tilde \alpha_1 + \tilde \alpha_4^u>0$ \cite{BurgosMarcos:2025xja}. The ``$\times$" represents the best fit point, while the QCDF result for $\alpha_1+\alpha_2$ from \cite{Beneke:2009ek} is overlaid for comparison.
    }
    \label{fig:alpha12}
\end{figure}
A natural question is how the hadronic parameters extracted from the factorizable $SU(3)_F$-breaking global fits compare to the QCDF predictions for the corresponding topological coefficients discussed in Sec.~\ref{sec:QCDF}. This comparison, however, is not straightforward: individual QCDF coefficients are largely unconstrained by the fit, for two main reasons:
\paragraph{Intrinsic blind directions.} Certain combinations of coefficients are intrinsically linked at the amplitude level, and are therefore individually unobservable regardless of the precision of the data available. This leads to a redundancy in the parametrization, without reflecting any theoretical assumption. For instance, the colour-allowed tree $\alpha_1$ and colour-suppressed EW penguin $\alpha_{4,\mathrm{EW}}^u$ always contribute together under the same CKM structure due to their corresponding operator structure. Therefore, global fits are sensitive only to the combination:
\begin{equation}
    \tilde \alpha_1 = \alpha_1 + \frac{3}{2}\alpha_{4,\mathrm{EW}}^u.
\end{equation}
Many other coefficients are affected by analogous blind directions \cite{BurgosMarcos:2025xja}.
\paragraph{Approximate blind directions.} In addition to this intrinsic redundancy, the current data is not yet precise enough to disentangle certain coefficient combinations that could, a priori, be determined separately. For example, all channels receiving contributions from $\tilde\alpha_1$, including $B^0\to\pi^+\pi^-, B_s^0\to K^-\pi^+, B^0\to K^+\pi^-$ and $B_s^0\to K^+K^-$, depend on it only through the combination $\tilde\alpha_1+\tilde\alpha_4^u$. While $\tilde\alpha_4^u$ can be constrained by $B^0_{(s)}\to K^0\bar K^0$, the current experimental uncertainties in these channels limit its determination and, consequently, that of $\tilde\alpha_1$. The fit saturates the experimental uncertainty of these channels such that this approximate blind direction remains largely unresolved with current measurements, reinforcing the point made in Sec.~\ref{sec:SU(3)sym} that improved precision on $B^0_{(s)}\to K^0\bar K^0$ would be particularly valuable.

Despite these limitations, comparisons with QCDF predictions are possible for combinations of coefficients that are both fit-constrained and available at higher perturbative order. This is the case for $\tilde \alpha_1+\tilde \alpha_2$, provided that EW penguin contributions are neglected. A comparison between the NNLO calculation for $\alpha_1(\pi\pi)$ and $\alpha_2(\pi\pi)$ \cite{Beneke:2009ek} and the results of the global fit including $SU(3)_F$ factorizable corrections \cite{BurgosMarcos:2025xja}, shown in Fig~\ref{fig:alpha12}, reveals a $2\sigma$ tension.

A similar strategy can be applied to probe the annihilation sector, where QCDF predictions are considerably more model-dependent due to the endpoint-divergent integrals discussed in Sec.~\ref{sec:challengeQCDF}. Phenomenological, data-driven analyses offer a way to constrain the size of these divergences directly. Measurements of CP asymmetries in the annihilation-dominated channels $B_s^0\to\pi^+\pi^-$ and $B^0\to K^+K^-$ would be particularly valuable, as they provide direct sensitivity to the size and phase of endpoint divergences.

\section{Conclusions}\label{sec:conclusions}
Calculating hadronic $B$ decays remains one of the major theoretical challenges in flavour physics, as these decays are largely dominated by non-perturbative QCD dynamics that are not yet reliably calculable from first principles, for example using lattice QCD.
QCD Factorization offers a framework for computing these hadronic topologies, order by order in $\alpha_s$ and at leading power in $\Lambda_\mathrm{QCD}/m_b$. Its main limitation lies in the endpoint divergences affecting hard spectator scattering and especially annihilation topologies, which are not calculable within the framework and require model-dependent phenomenological input.

Flavour symmetries provide a complementary way to study hadronic decays by relating the topological amplitudes of otherwise unrelated decay channels. These relations enable data-driven phenomenological analyses to constrain these amplitudes directly from experimental measurements. However, exact $SU(3)_F$ symmetry is not sufficient to describe $B\to M_1M_2$ decays. Once factorizable $SU(3)_F$-breaking corrections are included, inspired by the structure of QCD Factorization, an excellent description of the data is achieved. This approach is not limited to describing existing measurements, but can also predict observables for which experimental data is not available and identify which channels would most benefit from improved experimental precision, guiding future measurements.

Altogether, the interplay between QCD Factorization and data-driven $SU(3)_F$ analyses is only beginning to be fully exploited: the increasing precision of upcoming experimental data, together with progress on both fronts, will allow these theoretical approaches to be tested with unprecedented scrutiny.

\bibliography{References}

@article{Beneke:2009ek,
    author = "Beneke, M. and Huber, T. and Li, Xin-Qiang",
    title = "{NNLO vertex corrections to non-leptonic $B$ decays: Tree amplitudes}",
    eprint = "0911.3655",
    archivePrefix = "arXiv",
    primaryClass = "hep-ph",
    reportNumber = "PITHA-09-28, SFB-CPP-09-106, SI-HEP-2009-17",
    doi = "10.1016/j.nuclphysb.2010.02.002",
    journal = "Nucl. Phys. B",
    volume = "832",
    pages = "109--151",
    year = "2010"
}

@article{Berthiaume:2023kmp,
    author = {Berthiaume, Rapha{\"e}l and Bhattacharya, Bhubanjyoti and Boumris, Rida and Jean, Alexandre and Kumbhakar, Suman and London, David},
    title = "{Anomalies in Hadronic $B$ Decays}",
    eprint = "2311.18011",
    archivePrefix = "arXiv",
    primaryClass = "hep-ph",
    reportNumber = "UdeM-GPP-TH-23-299",
    doi = "10.1103/PhysRevLett.133.211802",
    journal = "Phys. Rev. Lett.",
    volume = "133",
    number = "21",
    pages = "211802",
    year = "2024"
}

@article{Bhattacharya:2025wcq,
    author = "Bhattacharya, Bhubanjyoti and Bouchard, Marianne and Hudy, Luke and Jean, Alexandre and London, David and MacKenzie, Christopher",
    title = "{Anomalies in hadronic B decays: An update}",
    eprint = "2505.11492",
    archivePrefix = "arXiv",
    primaryClass = "hep-ph",
    reportNumber = "UdeM-GPP-TH-25-306",
    doi = "10.1103/5jp1-7cfw",
    journal = "Phys. Rev. D",
    volume = "112",
    number = "5",
    pages = "056014",
    year = "2025"
}

@article{BurgosMarcos:2025xja,
    author = "Burgos Marcos, M. and Reboud, M. and Vos, K. K.",
    title = "{Detailed $SU(3)$ Flavour Symmetry Analysis of Charmless Two-Body $B$-Meson Decays Including Factorizable Corrections}",
    eprint = "2504.05209",
    archivePrefix = "arXiv",
    primaryClass = "hep-ph",
    reportNumber = "EOS-2025-01, Nikhef 2025-004",
    month = "4",
    year = "2025"
}

@article{Beneke:2003zv,
    author = "Beneke, Martin and Neubert, Matthias",
    title = "{QCD factorization for $B \rightarrow PP$ and $B \rightarrow PV$ decays}",
    eprint = "hep-ph/0308039",
    archivePrefix = "arXiv",
    reportNumber = "CLNS-03-1835, PITHA-03-06",
    doi = "10.1016/j.nuclphysb.2003.09.026",
    journal = "Nucl. Phys. B",
    volume = "675",
    pages = "333--415",
    year = "2003"
}

@article{PDG2025,
  author = "{Particle Data Group}",
  title = "{Review of Particle Physics}",
  journal = "Prog. Theor. Exp. Phys.",
  volume = "2025",
  year = "2025",
  note = "Section on $B^0$--$\bar B^0$ Mixing",
  url = "https://pdg.lbl.gov/2025/web/viewer.html?file=https://pdg.lbl.gov/2025/reviews/rpp2025-rev-b-bar-mixing.pdf"
}

@article{Huber:2021cgk,
    author = "Huber, Tobias and Tetlalmatzi-Xolocotzi, Gilberto",
    title = "{Estimating QCD-factorization amplitudes through SU(3) symmetry in $B\rightarrow P P$ decays}",
    eprint = "2111.06418",
    archivePrefix = "arXiv",
    primaryClass = "hep-ph",
    reportNumber = "SI-HEP-2021-31, P3H-21-092",
    doi = "10.1140/epjc/s10052-022-10068-8",
    journal = "Eur. Phys. J. C",
    volume = "82",
    number = "3",
    pages = "210",
    year = "2022"
}

@article{Beneke:1999br,
    author = "Beneke, M. and Buchalla, G. and Neubert, M. and Sachrajda, Christopher T.",
    title = "{QCD factorization for $B \rightarrow \pi \pi$ decays: Strong phases and CP violation in the heavy quark limit}",
    eprint = "hep-ph/9905312",
    archivePrefix = "arXiv",
    reportNumber = "SLAC-PUB-8146, CERN-TH-99-126, SHEP-99-04",
    doi = "10.1103/PhysRevLett.83.1914",
    journal = "Phys. Rev. Lett.",
    volume = "83",
    pages = "1914--1917",
    year = "1999"
}

@article{Beneke:2000ry,
    author = "Beneke, M. and Buchalla, G. and Neubert, M. and Sachrajda, Christopher T.",
    title = "{QCD factorization for exclusive, nonleptonic B meson decays: General arguments and the case of heavy light final states}",
    eprint = "hep-ph/0006124",
    archivePrefix = "arXiv",
    reportNumber = "CERN-TH-2000-159, CLNS-00-1675, PITHA-00-06, SHEP-00-06",
    doi = "10.1016/S0550-3213(00)00559-9",
    journal = "Nucl. Phys. B",
    volume = "591",
    pages = "313--418",
    year = "2000"
}

@article{FermilabLattice:2015mwy,
    author = "Bailey, Jon A. and others",
    collaboration = "Fermilab Lattice, MILC",
    title = "{$|V_{ub}|$ from $B\to\pi\ell\nu$ decays and (2+1)-flavor lattice QCD}",
    eprint = "1503.07839",
    archivePrefix = "arXiv",
    primaryClass = "hep-lat",
    reportNumber = "FERMILAB-PUB-15-108-T",
    doi = "10.1103/PhysRevD.92.014024",
    journal = "Phys. Rev. D",
    volume = "92",
    number = "1",
    pages = "014024",
    year = "2015"
}

@article{Flynn:2015mha,
    author = "Flynn, J. M. and Izubuchi, T. and Kawanai, T. and Lehner, C. and Soni, A. and Van de Water, R. S. and Witzel, O.",
    title = "{$B \to \pi \ell \nu$ and $B_s \to K \ell \nu$ form factors and $|V_{ub}|$ from 2+1-flavor lattice QCD with domain-wall light quarks and relativistic heavy quarks}",
    eprint = "1501.05373",
    archivePrefix = "arXiv",
    primaryClass = "hep-lat",
    reportNumber = "FERMILAB-PUB-15-019-T",
    doi = "10.1103/PhysRevD.91.074510",
    journal = "Phys. Rev. D",
    volume = "91",
    number = "7",
    pages = "074510",
    year = "2015"
}

@article{Colquhoun:2022atw,
    author = "Colquhoun, Brian and Hashimoto, Shoji and Kaneko, Takashi and Koponen, Jonna",
    collaboration = "JLQCD",
    title = {{Form factors of B{\textrightarrow}{\ensuremath{\pi}}{\ensuremath{\ell}}{\ensuremath{\nu}} and a determination of $|Vub|$ with M{\"o}bius domain-wall fermions}},
    eprint = "2203.04938",
    archivePrefix = "arXiv",
    primaryClass = "hep-lat",
    reportNumber = "KEK-CP-382",
    doi = "10.1103/PhysRevD.106.054502",
    journal = "Phys. Rev. D",
    volume = "106",
    number = "5",
    pages = "054502",
    year = "2022"
}

@article{Bouchard:2014ypa,
    author = "Bouchard, C. M. and Lepage, G. Peter and Monahan, Christopher and Na, Heechang and Shigemitsu, Junko",
    title = "{$B_s \to K \ell \nu$ form factors from lattice QCD}",
    eprint = "1406.2279",
    archivePrefix = "arXiv",
    primaryClass = "hep-lat",
    doi = "10.1103/PhysRevD.90.054506",
    journal = "Phys. Rev. D",
    volume = "90",
    pages = "054506",
    year = "2014"
}

@article{Gubernari:2018wyi,
    author = "Gubernari, Nico and Kokulu, Ahmet and van Dyk, Danny",
    title = "{$B\to P$ and $B\to V$ Form Factors from $B$-Meson Light-Cone Sum Rules beyond Leading Twist}",
    eprint = "1811.00983",
    archivePrefix = "arXiv",
    primaryClass = "hep-ph",
    reportNumber = "EOS-2018-02, TUM-HEP-1172/18",
    doi = "10.1007/JHEP01(2019)150",
    journal = "JHEP",
    volume = "01",
    pages = "150",
    year = "2019"
}

@article{Melic:2025uha,
    author = "Meli{\'c}, Bla{\v{z}}enka and Reboud, M{\'e}ril",
    title = "{Form factors and phenomenology of B$_{(s)}$ and D$_{(s)}$ semileptonic decays to {\ensuremath{\eta}} and $\eta'$}",
    eprint = "2507.13734",
    archivePrefix = "arXiv",
    primaryClass = "hep-ph",
    reportNumber = "RBI-ThPhys-2025-29, EOS-2025-04",
    doi = "10.1007/JHEP10(2025)025",
    journal = "JHEP",
    volume = "10",
    pages = "025",
    year = "2025"
}

@article{Leljak:2021vte,
    author = "Leljak, Domagoj and Meli{\'c}, Bla{\v{z}}enka and van Dyk, Danny",
    title = "{The $ \overline{B} $ {\textrightarrow} {\ensuremath{\pi}} form factors from QCD and their impact on $|V_{ub}|$}",
    eprint = "2102.07233",
    archivePrefix = "arXiv",
    primaryClass = "hep-ph",
    reportNumber = "EOS-2021-02, TUM-HEP-1316/21, RBI-ThPhys-2021-1",
    doi = "10.1007/JHEP07(2021)036",
    journal = "JHEP",
    volume = "07",
    pages = "036",
    year = "2021"
}

@article{LHCb:2008vvz,
    author = "Alves, Jr., A. Augusto and others",
    collaboration = "LHCb",
    title = "{The LHCb Detector at the LHC}",
    reportNumber = "LHCb-DP-2008-001",
    doi = "10.1088/1748-0221/3/08/S08005",
    journal = "JINST",
    volume = "3",
    pages = "S08005",
    year = "2008"
}

@article{Belle-II:2018jsg,
    author = "Altmannshofer, W. and others",
    editor = "Kou, E. and Urquijo, P.",
    collaboration = "Belle-II",
    title = "{The Belle II Physics Book}",
    eprint = "1808.10567",
    archivePrefix = "arXiv",
    primaryClass = "hep-ex",
    reportNumber = "KEK Preprint 2018-27, BELLE2-PUB-PH-2018-001, FERMILAB-PUB-18-398-T, JLAB-THY-18-2780, INT-PUB-18-047, UWThPh 2018-26",
    doi = "10.1093/ptep/ptz106",
    journal = "PTEP",
    volume = "2019",
    number = "12",
    pages = "123C01",
    year = "2019",
    note = "[Erratum: PTEP 2020, 029201 (2020)]"
}

@article{Li:2012cfa,
    author = "Li, Hsiang-nan and Lu, Cai-Dian and Yu, Fu-Sheng",
    title = "{Branching ratios and direct CP asymmetries in $D\to PP$ decays}",
    eprint = "1203.3120",
    archivePrefix = "arXiv",
    primaryClass = "hep-ph",
    doi = "10.1103/PhysRevD.86.036012",
    journal = "Phys. Rev. D",
    volume = "86",
    pages = "036012",
    year = "2012"
}

@article{Qin:2013tje,
    author = {Qin, Qin and Li, Hsiang-nan and L{\"u}, Cai-Dian and Yu, Fu-Sheng},
    title = "{Branching ratios and direct CP asymmetries in $D\to PV$ decays}",
    eprint = "1305.7021",
    archivePrefix = "arXiv",
    primaryClass = "hep-ph",
    doi = "10.1103/PhysRevD.89.054006",
    journal = "Phys. Rev. D",
    volume = "89",
    number = "5",
    pages = "054006",
    year = "2014"
}

@article{Zhou:2015jba,
    author = "Zhou, Si-Hong and Wei, Yan-Bing and Qin, Qin and Li, Ying and Yu, Fu-Sheng and Lu, Cai-Dian",
    title = "{Analysis of Two-body Charmed $B$ Meson Decays in Factorization-Assisted Topological-Amplitude Approach}",
    eprint = "1509.04060",
    archivePrefix = "arXiv",
    primaryClass = "hep-ph",
    doi = "10.1103/PhysRevD.92.094016",
    journal = "Phys. Rev. D",
    volume = "92",
    number = "9",
    pages = "094016",
    year = "2015"
}

@article{Beneke:2001ev,
    author = "Beneke, M. and Buchalla, G. and Neubert, M. and Sachrajda, Christopher T.",
    title = "{QCD factorization in $B \to \pi K, \pi \pi$ decays and extraction of Wolfenstein parameters}",
    eprint = "hep-ph/0104110",
    archivePrefix = "arXiv",
    reportNumber = "CERN-TH-2001-107, CLNS-01-1728, PITHA-01-01, SHEP-01-11",
    doi = "10.1016/S0550-3213(01)00251-6",
    journal = "Nucl. Phys. B",
    volume = "606",
    pages = "245--321",
    year = "2001"
}

@article{He:2018php,
    author = "He, Xiao-Gang and Wang, Wei",
    title = "{Flavor SU(3) Topological Diagram and Irreducible Representation Amplitudes for Heavy Meson Charmless Hadronic Decays: Mismatch and Equivalence}",
    eprint = "1803.04227",
    archivePrefix = "arXiv",
    primaryClass = "hep-ph",
    doi = "10.1088/1674-1137/42/10/103108",
    journal = "Chin. Phys. C",
    volume = "42",
    number = "10",
    pages = "103108",
    year = "2018"
}

@article{Bell:2015koa,
    author = "Bell, Guido and Beneke, Martin and Huber, Tobias and Li, Xin-Qiang",
    title = "{Two-loop current{\textendash}current operator contribution to the non-leptonic QCD penguin amplitude}",
    eprint = "1507.03700",
    archivePrefix = "arXiv",
    primaryClass = "hep-ph",
    reportNumber = "OUTP-15-15P, TUM-HEP-1005-15, SI-HEP-2015-17, QFET-2015-23",
    doi = "10.1016/j.physletb.2015.09.037",
    journal = "Phys. Lett. B",
    volume = "750",
    pages = "348--355",
    year = "2015"
}

@article{Bordone:2020gao,
    author = "Bordone, Marzia and Gubernari, Nico and Huber, Tobias and Jung, Martin and van Dyk, Danny",
    title = "{A puzzle in $\bar{B}_{(s)}^0 \to D_{(s)}^{(*)+} \lbrace \pi^-, K^-\rbrace$ decays and extraction of the $f_s/f_d$ fragmentation fraction}",
    eprint = "2007.10338",
    archivePrefix = "arXiv",
    primaryClass = "hep-ph",
    reportNumber = "TUM-HEP 1271/20, P3H-20-034, SI-HEP-2020-17",
    doi = "10.1140/epjc/s10052-020-08512-8",
    journal = "Eur. Phys. J. C",
    volume = "80",
    number = "10",
    pages = "951",
    year = "2020"
}

@article{Fleischer:2018bld,
    author = "Fleischer, Robert and Jaarsma, Ruben and Malami, Eleftheria and Vos, K. Keri",
    title = "{Exploring $B\rightarrow \pi \pi , \pi K$ decays at the high-precision frontier}",
    eprint = "1806.08783",
    archivePrefix = "arXiv",
    primaryClass = "hep-ph",
    reportNumber = "Nikhef-2018-021, SI-HEP-2018-18, QFET-2018-11, NIKHEF-2018-021",
    doi = "10.1140/epjc/s10052-018-6397-5",
    journal = "Eur. Phys. J. C",
    volume = "78",
    number = "11",
    pages = "943",
    year = "2018"
}

@article{Gronau:1990ka,
    author = "Gronau, Michael and London, David",
    title = "{Isospin analysis of CP asymmetries in B decays}",
    reportNumber = "DESY-90-106-REV, UDEM-LPN-TH24-REV",
    doi = "10.1103/PhysRevLett.65.3381",
    journal = "Phys. Rev. Lett.",
    volume = "65",
    pages = "3381--3384",
    year = "1990"
}

@article{Gronau:2005pq,
    author = "Gronau, M. and Zupan, J.",
    title = "{Isospin-breaking effects on alpha extracted in $B \to \pi \pi, \rho \rho, \rho \pi$}",
    eprint = "hep-ph/0502139",
    archivePrefix = "arXiv",
    reportNumber = "TECHNION-PH-2005-3",
    doi = "10.1103/PhysRevD.71.074017",
    journal = "Phys. Rev. D",
    volume = "71",
    pages = "074017",
    year = "2005"
}

@article{Belle-II:2026kzq,
    author = "Abumusabh, Merna and others",
    collaboration = "Belle-II",
    title = "{Measurement of mixing-induced CP violation in the decay $B^0 \to \pi^0 \pi^0$}",
    eprint = "2606.20797",
    archivePrefix = "arXiv",
    primaryClass = "hep-ex",
    reportNumber = "Belle II Preprint 2026-013, KEK Preprint 2026-9",
    month = "6",
    year = "2026"
}

@article{Gronau:2000zy,
    author = "Gronau, Michael",
    title = "{U-spin symmetry in charmless B decays}",
    eprint = "hep-ph/0008292",
    archivePrefix = "arXiv",
    reportNumber = "CERN-TH-2000-250",
    doi = "10.1016/S0370-2693(00)01119-9",
    journal = "Phys. Lett. B",
    volume = "492",
    pages = "297--302",
    year = "2000"
}

@article{Gronau:2013mda,
    author = "Gronau, Michael",
    title = "{U-spin breaking in CP asymmetries in B decays}",
    eprint = "1308.3448",
    archivePrefix = "arXiv",
    primaryClass = "hep-ph",
    doi = "10.1016/j.physletb.2013.09.056",
    journal = "Phys. Lett. B",
    volume = "727",
    pages = "136--140",
    year = "2013"
}

@article{Fleischer:1999pa,
    author = "Fleischer, Robert",
    title = "{New strategies to extract Beta and gamma from $B_{(d)} \to \pi+ \pi-$ and $B_{(s)} \to K+ K-$}",
    eprint = "hep-ph/9903456",
    archivePrefix = "arXiv",
    reportNumber = "CERN-TH-99-79",
    doi = "10.1016/S0370-2693(99)00640-1",
    journal = "Phys. Lett. B",
    volume = "459",
    pages = "306--320",
    year = "1999"
}

@article{Fang:2026fhl,
    author = "Fang, Wen-Sheng and Huber, Tobias and Li, Xin-Qiang and Malami, Eleftheria and Tetlalmatzi-Xolocotzi, Gilberto",
    title = "{QCD-factorization amplitudes from flavour symmetries: beyond the $SU(3)$ symmetric case}",
    eprint = "2604.19612",
    archivePrefix = "arXiv",
    primaryClass = "hep-ph",
    reportNumber = "SI-HEP-2026-09, P3H-26-029",
    month = "4",
    year = "2026"
}

@article{Bell:2020qus,
    author = "Bell, Guido and Beneke, Martin and Huber, Tobias and Li, Xin-Qiang",
    title = "{Two-loop non-leptonic penguin amplitude in QCD factorization}",
    eprint = "2002.03262",
    archivePrefix = "arXiv",
    primaryClass = "hep-ph",
    reportNumber = "TUM-HEP-1250/20, SI-HEP-2019-17, QFET-2019-12",
    doi = "10.1007/JHEP04(2020)055",
    journal = "JHEP",
    volume = "04",
    pages = "055",
    year = "2020"
}

@article{Keum:2000ph,
    author = "Keum, Yong-Yeon and Li, Hsiang-nan and Sanda, A. I.",
    title = "{Fat penguins and imaginary penguins in perturbative QCD}",
    eprint = "hep-ph/0004004",
    archivePrefix = "arXiv",
    reportNumber = "KEK-TH-642, NCKU-HEP-00-01, KIAS-P00014, DPNU-00-13",
    doi = "10.1016/S0370-2693(01)00247-7",
    journal = "Phys. Lett. B",
    volume = "504",
    pages = "6--14",
    year = "2001"
}

@article{Keum:2000wi,
    author = "Keum, Y. Y. and Li, Hsiang-Nan and Sanda, A. I.",
    title = "{Penguin enhancement and $B \to K \pi$ decays in perturbative QCD}",
    eprint = "hep-ph/0004173",
    archivePrefix = "arXiv",
    reportNumber = "NCKU-HEP-00-01A, APCTP-00-05, DPNU-00-14",
    doi = "10.1103/PhysRevD.63.054008",
    journal = "Phys. Rev. D",
    volume = "63",
    pages = "054008",
    year = "2001"
}

@article{Lu:2000em,
    author = "Lu, Cai-Dian and Ukai, Kazumasa and Yang, Mao-Zhi",
    title = "{Branching ratio and CP violation of $B\to \pi\pi$ decays in perturbative QCD approach}",
    eprint = "hep-ph/0004213",
    archivePrefix = "arXiv",
    reportNumber = "HUPD-9924, DPNU-00-15",
    doi = "10.1103/PhysRevD.63.074009",
    journal = "Phys. Rev. D",
    volume = "63",
    pages = "074009",
    year = "2001"
}

@article{Huber:2016xod,
    author = {Huber, Tobias and Kr{\"a}nkl, Susanne and Li, Xin-Qiang},
    title = "{Two-body non-leptonic heavy-to-heavy decays at NNLO in QCD factorization}",
    eprint = "1606.02888",
    archivePrefix = "arXiv",
    primaryClass = "hep-ph",
    reportNumber = "SI-HEP-2016-12, QFET-2016-06",
    doi = "10.1007/JHEP09(2016)112",
    journal = "JHEP",
    volume = "09",
    pages = "112",
    year = "2016"
}

@article{Beneke:2000pw,
    author = "Beneke, M.",
    editor = "Haber, Howard E.",
    title = "{Conceptual aspects of QCD factorization in hadronic B decays}",
    eprint = "hep-ph/0009328",
    archivePrefix = "arXiv",
    reportNumber = "PITHA-00-23",
    doi = "10.1088/0954-3899/27/6/303",
    journal = "J. Phys. G",
    volume = "27",
    pages = "1069--1080",
    year = "2001"
}
\end{document}